\documentclass[iop,apj]{emulateapj}

\shorttitle{Large-Scale Magnetic Fields in Accretion Disks}
\shortauthors{Takeuchi \& Okuzumi}

\begin{document}

\title{Radial Transport of Large-Scale Magnetic Fields in Accretion
Disks. II. Relaxation to Steady States}
\author{Taku Takeuchi\altaffilmark{1} and Satoshi Okuzumi}
\affil{Department of Earth and Planetary Sciences, Tokyo Institute of Technology, Meguro-ku, Tokyo, 152-8551, Japan}

\begin{abstract}
We study the time evolution of a large-scale magnetic flux threading
an accretion disk. Induction equation of the mean poloidal field is solved
under the standard viscous disk model. Magnetic flux evolution is
 controlled by the two timescales: One is the timescale of the inward
 advection of the magnetic flux, $\tau_{adv}$. This is induced by the
 dragging of the flux by the accreting gas. The other is the outward diffusion
 timescale of the magnetic flux $\tau_{dif}$. We consider diffusion due
 to the Ohmic resistivity. These timescales can be significantly
 different from the disk viscous timescale $\tau_{disk}$.
The behaviors of the magnetic flux evolution is quite different
depending on the magnitude relationship of the timescales $\tau_{adv}$,
 $\tau_{dif}$, and $\tau_{disk}$. The most interesting phenomena occurs
 when $\tau_{adv} \ll \tau_{dif}, \ \tau_{disk}$. In such a case, the
 magnetic flux distribution approaches a quasi-steady profile much
 faster than the viscous evolution of the gas disk, and also the magnetic
 flux has been tightly bundled to the inner part of the disk. In the inner
 part, although the poloidal magnetic field becomes much stronger than the
interstellar magnetic field, the field strength is limited to the maximum
 value that is analytically given by our previous work (Okuzumi et
 al. 2014, ApJ, 785, 127).
We also find a condition for that the initial large magnetic flux, which is a
 fossil of the magnetic field dragging during the early phase of star
 formation, survives for a duration in which significant gas disk
 evolution proceeds.

\end{abstract}


\altaffiltext{1}{taku@geo.titech.ac.jp}

\setcounter{footnote}{2}

\section{Introduction}
\label{sec:Introduction}

Magnetic field is an important ingredient of the evolutionary process of
accretion disks. A magnetic flux vertically threading a disk induces
disk accretion via magneto-rotational instability (MRI; Balbus \& Hawley
1998 for a review).  It also accelerates disk winds via the
magneto-centrifugal force (Blandford \& Payne 1982) or the magnetic
pressure (Shibata \& Uchida 1985). The activity of these processes
depends on the strength of the large scale magnetic field. For this reason,
how the strength of large scale field is determined has been a key
question in disk accretion processes. 

A simple model on evolution of a large scale magnetic field threading an
accretion disk was proposed by Lubow et al. (1994; henceforth
LPP94). This model solves the evolution of the mean poloidal field, which is
determined by the balance between inward dragging by the accreting
gas and outward diffusion of the magnetic flux. Although this model
neglects several important factors including the effect of the toroidal
field, the analyses of this model have been providing us a useful guide
in understanding the basic properties of the field transport in
accretion disks (e.g., Shu et al. 2007; Cao 2011; Cao \& Spruit 2013;
Guilet \& Ogilvie 2014).

Okuzumi et al. (2014; henceforth Paper I) have performed a comprehensive
analysis on steady field profiles under the LPP94 model. They
derived the maximum strength of the steady field, which is determined by
the external field strength and the disk size (see Figure 7 of
Paper I and see also Figure 9 of Guilet \& Ogilvie 2014). This maximum
field strength implies an upper limit on the accretion rate of disks due
to MRI(Hawley et al. 1995; Suzuki et al. 2010; Okuzumi \& Hirose 2011;
Gressel et al. 2012; Simon et al. 2013) and the mass loss rate due to
magnetically driven wind (Blandford \& Payne 1982; Bai \& Stone 2013;
Simon et al. 2013).
The results of Paper I and Guilet \& Ogilvie (2014) are thus providing
significant predictions on the accretion and wind mass loss rates
driven by the large scale magnetic field.

However, the analysis in Paper I was limited to steady state where
inward field dragging balances with outward diffusion of the magnetic
flux. Thus, the maximum field strength derived in Paper I is applicable
only after the initial field has relaxed to a quasi-steady
configuration. Before that, there remained the magnetic flux that had
been dragged from the parent cloud core. This paper is devoted to
analyzing how large-scale fields in disks evolve before the quasi-steady state
has been reached.

Here, we briefly summarize the timescale argument in the LPP94 model.
The relaxation timescale of mean poloidal fields is $r h/\eta$,
where $r$ the disk radius, $h$ the disk thickness, and $\eta$ is the magnetic
diffusivity (Equation (\ref{eq:tau_dif}); see also Section 1 of Lovelace
et. al. 2009). This is shorter than the disk evolution timescale, $r^2 /
\nu$, where $\nu$ is the gas viscosity, provided that the magnetic
Prandtl number $P_m = \nu /\eta$ is of order unity. Thus, quasi-steady
fields in the disks are expected. LPP94 has shown, however, that the advection
timescale of fields due to the drag by the accreting gas is also
estimated as $r^2 / \nu$, meaning that inward dragging of the field is
too weak to keep the magnetic flux against outward diffusion.
The above discussion predicts that the initial field configuration
quickly relaxes to a steady state. However, this argument also predicts
that effective inward transport of the magnetic flux would not occur.

The above estimate of the timescales is based on the assumption that the
field advection takes place at the disk equatorial plane and the
advection velocity of the field is similar to that of the gas. Ogilvie
\& Livio (2001) pointed out, however, that the gas drags the field
mainly at a certain height where the electric conductivity is
high. The advection velocity of the field would differ from
the advection velocity of the gas at the midplane (Rothstein et
al. 2008). If the field advection is much faster than the gas
advection, it may be possible that the field relaxes to a
quasi-steady configuration, in which a strong magnetic flux is still
maintained in the disk. Guilet \& Ogilvie (2014) has shown that such a
situation is realized when the accretion velocity at the disk surface is
much faster than that at the disk midplane. The vertical profile of the
accretion velocity is yet unclear, and several papers have been devoted
to studying the vertical structure of the disk (Lovelace et al. 2009;
Bisnovatyi-Kogan \& Lovelace 2013; Guilet \& Ogilvie 2012, 2013).

This paper is the subsequent paper of Paper I, which
has solved steady magnetic fields using the LPP94 model. The present
paper studies the time evolution of the magnetic fields. We assume that
the advection velocities of the gas and of the field are  different. How
much these velocities differ from each other depends on the vertical
structure of the disk. Guilet \& Ogilvie (2012, 2013) argued the
possibility that the field advection velocity is more than 10 times
greater than the gas advection. In this paper, we treat this factor,
$C_u$, as a free parameter and solve the radial profile of the magnetic
field. The vertically averaged magnetic Prandtl number, $P_{m,eff}$, is
also treated as a free parameter.  

The main purpose of this paper is to quantify the condition that the
magnetic field relaxes to a steady state faster than the gas evolution,
and that the field advection also occurs fast enough to keep a
significant magnetic flux in the disk. We derive basic equations in
Section \ref{sec:Equations}. The main analytical result on the above
condition is expressed on the $P_{m,eff}$-$C_u$ plane in Section
\ref{sec:timescales}, and is 
numerically verified in Section \ref{sec:Bz_evol}. If this condition is
satisfied in realistic accretion disks, the maximum field strength in
disks is limited by the prediction proposed in Paper I. The results are
discussed in Section \ref{sec:Discussions} and summarized in Section
\ref{sec:summary}.

\section{Basic Equations}
\label{sec:Equations}

\subsection{A Gas Disk}
\label{sec:GasEqs}

We consider an accretion disk which evolves via the
``$\alpha$-viscosity''. The kinematic viscosity is written 
as $\nu=\alpha c_s h$, where $c_s$ is the sound speed, $h$ is the
half thickness of the disk, and the parameter $\alpha$ is constant 
throughout the disk. The disk is assumed to be geometrically thin, with
the ratio
\begin{equation}
\varepsilon = \frac{h}{r} \ 
\label{eq:varepsiron}
\end{equation}
much smaller than unity. Using this prescription and the Keplerian rotation
profile $\Omega(r) \propto r^{-3/2}$, the accretion velocity of the gas
is written as (e.g., Frank et al. 1992)
\begin{equation}
\bar{u_r} = - \frac{3}{r^{1/2} \Sigma} \frac{\partial}{\partial r}
 \left( r^{1/2} \bar{\nu} \Sigma \right) \ . 
\label{eq:ur}
\end{equation}
Here $\bar{u_r}$ and $\bar{\nu}$ are mass-weighted vertical averages of
the radial velocity and viscosity, respectively:
\begin{equation}
\bar{u_r}=\frac{\int_{-h}^{h} \rho u_r dz}{\int_{-h}^{h}
  \rho dz} \ , \ \ \ 
\bar{\nu}=\frac{\int_{-h}^{h} \rho \nu dz}{\int_{-h}^{h}
  \rho dz} \ ,
\label{eq:ur_denave}
\end{equation}
where $\rho$ is the mass density of the gas. The evolution of the disk surface density is determined by 
\begin{equation}
\frac{\partial\Sigma}{\partial t}-\frac{3}{r}\frac{\partial}{\partial
r}\left[  r^{1/2}\frac{\partial}{\partial r}\left(  r^{1/2} \bar{\nu}
  \Sigma\right) \right]  =0 \ .
\label{eq:sigma_evol}
\end{equation}

\begin{figure}[ptb]
\epsscale{1.2}
\plotone{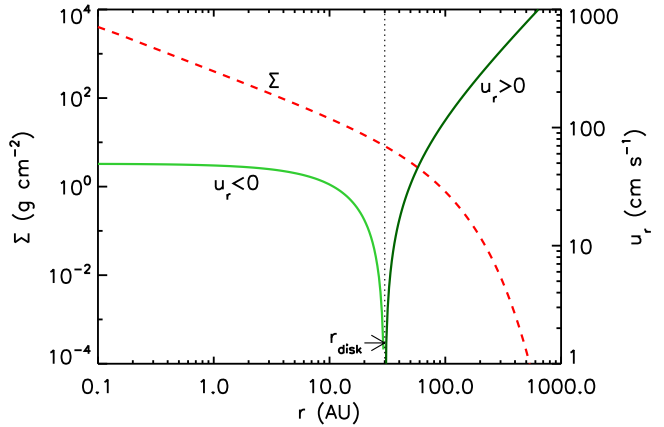}
\caption{Structure of the gas disk. The red dashed line shows the
 surface density profile $\Sigma$. The green solid line shows the radial
 velocity $u_r$. The radial velocity is negative for $r < r_{disk}$ and
 positive for $r > r_{disk}$.} 
\label{fig:gas_similar}
\end{figure}

Equation (\ref{eq:sigma_evol}) has similarity solutions, when
the viscosity profile obeys a power-law form,
$\bar{\nu}=\bar{\nu_0}(r/r_0)^{\gamma}$ 
(Lynden-Bell \& Pringle 1974; Hartmann et al. 1998; Kitamura et
al. 2002). The similarity solution is  written as
\begin{equation}
 \Sigma(r,t)  = {\Sigma _0}{\left( {\frac{r}{{{r_0}}}} \right)^{ - \gamma
 }}{T_g}^{ - \frac{{5 - 2\gamma }}{{4 - 2\gamma }}}\exp \left[ { -
 \frac{1}{{2(2 - \gamma )}}{{\left( {\frac{r}{{{r_{disk}}}}} \right)}^{2
 - \gamma }}} \right] \ ,
\end{equation}
and
\begin{equation}
\bar{u_r}(r,t) =  - \frac{{3 \bar{\nu_0}}}{{2r}}{\left( {\frac{r}{{{r_0}}}}
\right)^\gamma }\left[ {1 - {{\left( {\frac{r}{{{r_{disk}}}}}
 \right)}^{2 - \gamma }}} \right] \ , 
\label{eq:ur_similar}
\end{equation}
where
\begin{equation}
T_g = \frac{{3(2 - \gamma ) \bar{{\nu _0}}}}{{2{r_0}^2}}t + 1 \ ,
\end{equation}
\begin{equation}
r_{disk}(t)=T_g^{1/(2-\gamma)} r_0 \ .
\end{equation}
Here $T_g$ is the non-dimensional time scaled by the viscous time at
$r_0$, and $r_{disk}$ is the disk size defined as such that the disk gas
accretes ($\bar{u_r}<0$) for $r<r_{disk}$ and it diffuses outward
($\bar{u_r}>0$) for $r>r_{disk}$ (see Equation (\ref{eq:ur_similar})).
We define $r_0$ as the disk size at $t=0$, i.e., $r_0=r_{disk}(t=0)$.
This similarity solution is shown in Figure \ref{fig:gas_similar} for
$r_{disk}=30$AU. 

We consider a protoplanetary disk around a young star of $1
M_{\sun}$ for numerical calculations. The temperature is assumed to
be $T \propto r^{-q}$. The following parameters are used for our
fiducial model,
\begin{equation}
\varepsilon  = 7.78 \times {10^{ - 2}}{\left( {\frac{r}{{{r_0}}}}
 \right)^{\frac{{1 - q}}{2}}} \ ,
\end{equation}
\begin{equation}
\bar{\nu}  = 1.48 \times {10^{16}}{\left( {\frac{r}{{{r_0}}}}
			     \right)^{\frac{3}{2} - q}}{\rm{
c}}{{\rm{m}}^2} \ {{\rm{s}}^{ - 1}} \ .
\end{equation}
In the numerical calculations, we adopt $r_0=30$AU and $q=1/2$. The power-law index of the viscosity is
$\gamma=3/2-q=1$. The numerical values are calculated for $T=278 (r/1 \
\rm{AU})^{-1/2}$ K and $\alpha=10^{-2}$. The initial surface density is
calculated from the mass accretion rate to the star, as $\Sigma_0
=\dot{M}/(3 \pi \nu_0)$. For $\dot{M}=3 \times 10^{-8} M_{\sun} \
\rm{yr}^{-1}$, $\Sigma_0=13.6 \ \rm{g \ cm}^{-2}$. (Note that $\Sigma$
does not appear in the induction equation (\ref{eq:induction_ave}) and
thus the above choice of $\Sigma_0$ does not affect the magnetic field
evolution.)

\subsection{Induction Equation}
\label{sec:InductionEqs}

Magnetic field evolution is solved using the model proposed by
LPP94. This model is the simplest model that describes the evolution of the
poloidal magnetic flux threading an accretion disk. The toroidal field
is neglected and the axisymmetric poloidal field is determined so as to
connect to an external uniform field at infinity.  

We consider evolution of the mean poloidal field threading a turbulent
disk. The mean field inside the disk is calculated by Reynolds
averaging of the turbulent field. Assuming that the mean poloidal field is
axisymmetric, it is written by a flux function $\psi(r,z,t)$ as
\begin{equation}
B_r =-\frac{1}{r}\frac{\partial \psi}{\partial z} \ ,  \ \ \ 
B_z =\frac{1}{r}\frac{\partial \psi}{\partial r} \ .
\label{eq:flux_func}
\end{equation}
The induction equation for the mean poloidal field is, as shown by LPP94,
\begin{equation}
\frac{{\partial \psi }}{{\partial t}} =  - {u_r}\frac{{\partial \psi
}}{{\partial r}} - r\eta \frac{4 \pi J_{\phi} }{c} \ ,
\label{eq:induction1}
\end{equation}
where $u_r$ is the mean radial velocity of the gas, $\eta$ is the
magnetic diffusivity, $J_{\phi}$ is the azimuthal current density, and
$c$ is the speed of light.

We average Equation (\ref{eq:induction1}) vertically in the following way. As
pointed out by Ogilvie \& Livio (2001), conductivity-weighted averaging
is the proper way for averaging. Multiplying Equation
(\ref{eq:induction1}) by the conductivity $\sigma_e = c^2 / (4 \pi
\eta)$ and integrating over $z$, we obtain
\begin{equation}
\frac{{\partial \psi }}{{\partial t}} =  - {u_{r*}}\frac{{\partial \psi
}}{{\partial r}} - \frac{r \eta_{*}}{2h} \frac{{4\pi {K_{\phi} }}}{c} \ ,
\label{eq:induction_ave}
\end{equation}
where $K_{\phi} =\int_{-h}^h J_\phi dz$ is the surface current
density, and the subscript ``$\ast$'' means the conductivity-weighted
averages, such that
\begin{equation}
u_{r*}=\frac{\int_{-h}^{h} \sigma_{e} u_r dz}{\int_{-h}^{h}
  \sigma_{e} dz} 
\label{eq:ur_condave}
\end{equation}
\begin{equation}
\eta_{*}=\frac{\int_{-h}^{h} \sigma_{e} \eta dz}{\int_{-h}^{h}
  \sigma_{e} dz} = \frac{h c^2} {2 \pi \int_{-h}^{h}
  \sigma_{e} dz} \ .
\label{eq:eta_condave}
\end{equation}
Here, the integration has been done from $-h$ to $h$, as in the
mass-weighted vertical averaging (Equation (\ref{eq:ur_denave})). This
is just a rough estimate. In realty, in the disk corona ($|z|>h_B$,
where $h_B$ is the height at which the magnetic 
pressure is same as the thermal pressure), the gas cannot drag the
magnetic field and the current vanishes under our assumption of
axisymmetric poloidal field. Thus, the integration in Equations
(\ref{eq:ur_condave}) and (\ref{eq:eta_condave}) should be done for $|z|
< h_B$, where the thermal pressure dominates the magnetic pressure, and
it should be noted that $h_B$ varies with the field strength. (See
Guilet \& Ogilvie (2012) for detailed discussion.)
The surface current of the disk $K_{\phi}$ is related to $\psi$ via the Biot-Savart equation,
\begin{equation}
\psi - \psi_{\infty}=\frac{r}{c}\int_{r_{in}}^{r_{out}}\int_{0}^{2\pi}\frac
{K_{\phi}(r^{\prime})\cos\phi^{\prime}d\phi^{\prime}r^{\prime}dr^{\prime}
}{\left(  r^{2}+r^{\prime2}-2rr^{\prime}\cos\phi^{\prime}\right)
  ^{1/2}} \ ,
\label{eq:BiotSavart}
\end{equation}
where $\psi_{\infty}$ is the flux of the external field. When the
constant external field $B_{\infty}$ is considered, we have
\begin{equation}
\psi_{\infty}=\frac{1}{2}B_{\infty}r^{2}~.
\end{equation}

Note that we have taken conductivity-weighted averaging for the induction equation, while mass-weighted averaging is used for the density
evolution. In principle, $u_{r*}$ is not necessary equal to
$\bar{u_r}$. Therefore, we introduce a ratio of the average accretion
velocities, $C_u$, as
\begin{equation}
u_{r*}=C_u \bar{u_r} \ .
\label{eq:ur_ave_ratio}
\end{equation}
The effective Prandtl number is defined by the ratio of the
mass-weighted average of the gas viscosity to the conductivity-weighted 
average of the magnetic diffusivity,
\begin{equation}
P_{m,eff} =\frac{\bar{\nu}}{\eta_{*}} \ .
\label{eq:Pmeff}
\end{equation}

\section{Timescale Argument and Quasi-steady States}
\label{sec:timescales}

From the equations for the gas and magnetic flux evolutions (Equations
(\ref{eq:sigma_evol}) and (\ref{eq:induction_ave})), we define in Section
\ref{sec:various_timescales} various
evolution timescales, i.e., viscous timescale $\tau_{disk}$, the
diffusion timescale of the magnetic flux $\tau_{dif}$, and the advection
timescale of the magnetic flux $\tau_{adv}$. In the following
definition, we neglect numerical factors of order unity.

In section \ref{sec:timescale_comparison}, we show that the behaviors of
the magnetic flux evolution is quite different depending on the
magnitude relationship of the timescales $\tau_{adv}$, $\tau_{dif}$, and
$\tau_{disk}$. The most interesting phenomena occurs  when $\tau_{adv}
\ll \tau_{dif}$, $\tau_{disk}$. In such a case, the  magnetic flux
distribution approaches a quasi-steady profile much faster than the
viscous evolution of the gas disk, and the magnetic flux has been tightly bundled to the
inner part of the disk. In section \ref{sec:steady_profiles}, we derive
the analytical flux profiles for the cases of $\tau_{adv} \ll \tau_{dif}$,
$\tau_{disk}$. These analytic expression was derived mainly in Paper
I. In Appendix \ref{sec:B_outward}, we expand the results of Paper I to
include the effect of outward motion of the outer part of the viscous
gas disk.

The argument in this section will be verified by the numerical
integration of the basic equations in Section \ref{sec:Bz_evol}.

\subsection{Various Evolution Timescales}
\label{sec:various_timescales}

The viscous evolution of the gas disk is determined by Equation
(\ref{eq:sigma_evol}). Its timescale is given by
\begin{equation}
\tau_{disk} \equiv \frac{r_{disk}^2}{\bar{\nu}} \ .
\label{eq:tau_disk}
\end{equation}

The evolution timescale of the mean field is determined by the induction
equation. The second term of the right hand side of Equation
(\ref{eq:induction_ave}) causes diffusion of the magnetic field. The 
timescale of diffusion is estimated as $\tau_{dif}\sim (h c \psi)/(r
\eta_{*} K_{\phi})$. From the Biot-Savart equation
(\ref{eq:BiotSavart}), $K_{\phi}/c \sim \psi / r^2$ (see also
Equations (29) of Paper I), and then the diffusion timescale
is given by
\begin{equation}
\tau_{dif} \equiv \varepsilon P_{m,eff} \tau_{disk} \ .
\label{eq:tau_dif}
\end{equation}
This means that the diffusion timescale of the magnetic field is shorter
than that of the gas evolution timescale, as long as $P_{m,eff} \le
\varepsilon^{-1}$. If $P_{m,eff} \sim 1$, as usually expected, the
magnetic field relaxes quickly to a quasi-steady state within the disk
evolution timescale.

The first term of the right hand side of Equation (\ref{eq:induction_ave})
represents the advection of the magnetic field. The 
timescale of advection is estimated as $\tau_{adv} \sim r /
u_{r*} $.  From Equation (\ref{eq:ur_ave_ratio}) and $
\bar{u_r} \sim \bar{\nu} / r$, the advection timescale is given by
\begin{equation}
\tau_{adv} \equiv C_u^{-1} \tau_{disk} \ .
\label{eq:tau_adv}
\end{equation}
Thus, if $C_u > 1$, the advection timescale of the magnetic field
is shorter than the gas evolution timescale.

\subsection{Comparison of Timescales}
\label{sec:timescale_comparison}

\begin{figure}[ptb]
\epsscale{1.2}
\plotone{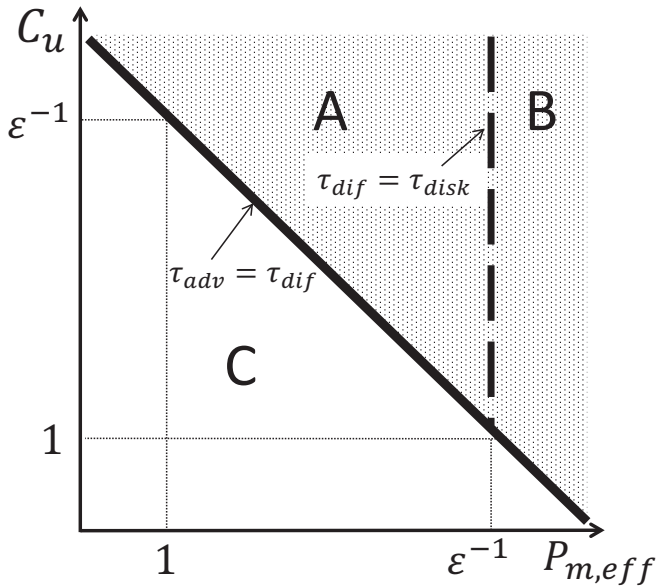}
\caption{Various evolution regimes on the $P_{m,eff}$-$C_u$ plane. 
The solid and dashed lines indicate the boundaries where $\tau_{adv} =
 \tau_{dif}$ and $\tau_{dif} = \tau_{disk}$, respectively.
In the shaded region ($\tau_{adv} < \tau_{dif}$), the effective field
 dragging by the gas takes place against the magnetic
 diffusion. This region is subdivided into regions A (where $\tau_{dif}
 < \tau_{disk}$) and B (where $\tau_{dif} > \tau_{disk}$).
In region A, the magnetic field relaxes to a quasi-steady state rapidly
 within the disk evolution timescale. In  region B, on the other hand,
 the initial profile of the magnetic field  at the disk formation stage
 remains during gas disk evolution. In region C (where $\tau_{adv} >
 \tau_{dif}$) magnetic diffusion dominates over field advection, and
 therefore any magnetic flux except for externally imposed flux,
 $\psi_{\infty}$, is lost. }
\label{fig:Pm-Cu}
\end{figure}

The condition for effective field advection to the inner part of the
disk is $\tau_{adv} < \tau_{dif}$. This condition reduces to 
\begin{equation}
C_u P_{m,eff} > \varepsilon^{-1} \ ,
\label{eq:cond_drag}
\end{equation}
and is shown by the shaded region on the $P_{m,eff}$-$C_u$ plane in
Figure \ref{fig:Pm-Cu}. 

Effective field advection ($\tau_{adv} \ll \tau_{dif}$) takes place
either if $C_u > \varepsilon^{-1}$ or if 
$P_{m,eff} > \varepsilon^{-1}$. First, we consider a case in which $C_u
> \varepsilon^{-1}$ and $P_{m,eff} \sim 1$. In this case, the magnetic
field relaxes to a quasi-steady state more quickly compared to the disk
evolution ($\tau_{dif} \ll \tau_{disk}$). Thus, the profile of the magnetic field at
any evolutionary stages should be given by the steady profile discussed in
Section \ref{sec:B_stdy} below, except at early stages. This regime
is shown as the ``region A'' in Figure \ref{fig:Pm-Cu}.

If the effective field advection is supported by large values of $P_{m,eff} >
\varepsilon^{-1}$ rather than by large $C_u$, then the magnetic
diffusion timescale $\tau_{dif} = P_{m,eff} \varepsilon \tau_{disk}$
is also longer than the disk evolution timescale $\tau_{disk}$. In such
cases, the magnetic field remembers its initial profile that was set at
the formation stage of the disk. This regime is shown as the
``region B'' in Figure \ref{fig:Pm-Cu}.

Finally, in the ``region C'' in Figure \ref{fig:Pm-Cu}, the condition
for significant field advection given by Equation (\ref{eq:cond_drag}) is
not satisfied. In such cases, disk accretion cannot drag the
magnetic field effectively. Even if a strong concentration of the magnetic
flux at the inner disk was created during the disk formation stage, the
magnetic flux diffuses outward with a timescale $\tau_{dif}$, which is
expected to be shorter than the disk evolution timescale $\tau_{disk}$ (as long as
$P_{m,eff} < \varepsilon^{-1}$). 

We stress that in the region A, i.e., if $\tau_{adv} \ll \tau_{dif} \ll
\tau_{disk}$, a large-scale magnetic field in an accretion disk has been
transferred to the inner part of the disk and relaxes into a
quasi-steady state. This interesting regime will be discussed in detail in the
next subsection.

\subsection{Quasi-Steady Profile of the Magnetic Field for Strong Dragging}
\label{sec:steady_profiles}

In this subsection, we focus on the region A in Figure \ref{fig:Pm-Cu}, in
which most of the magnetic flux that initially threaded the disk have
been transported to the inner part of the disk. We derive the
quasi-steady profiles of the magnetic field under given density and
velocity profiles of the gas disk. Setting $\partial / \partial t = 0$,
Equation (\ref{eq:induction_ave}) becomes 
\begin{equation}
B_z - \frac{D}{2} \frac{{4\pi {K_{\phi}}}}{c}=0 \ ,
\label{eq:induction_stdy}
\end{equation}
where
\begin{equation}
D=-\frac{\eta_{*}}{u_{r*} h} = - \frac{1}{P_{m,eff} C_u
 \varepsilon} \frac{\bar{\nu}}{\bar{u_r} r} \ ,
\label{eq:D_eff}
\end{equation}
denotes the effective diffusivity compared to the advection. For disks
with smooth density and velocity profiles, $\bar{\nu} / (\bar{u_r}
r) \sim 1$. Thus the magnitude of $D$ is roughly estimated as
\begin{equation}
| D | \sim (P_{m,eff} C_u \varepsilon)^{-1} =
 \frac{\tau_{adv}}{\tau_{dif}} \ .
\label{eq:D_mag}
\end{equation}
For significant field advection, $|D| \ll 1$ is required. 

\subsection{A Toy Model for a Gas Disk}
\label{sec:disk_struc}

As shown in Figure \ref{fig:gas_similar}, the inner part of the disk
($r<r_{disk}$) accretes to the star and the outer part diffuses
outward. The magnetic field is dragged to the same direction of this gas
motion. In the outer part, the gas density declines exponentially with
the radius, and at the outer edge it becomes so tenuous that the gas
pressure becomes lower than the magnetic pressure or that the ambipolar
diffusion suppresses the MRI (Walsh et al. 2012; Dzyurkevich et
al. 2013). Thus, at the outermost part, the gas cannot drag the magnetic field
effectively. This region is modeled by a large effective diffusivity $D$. 

\begin{figure}[ptb]
\epsscale{1.2}
\plotone{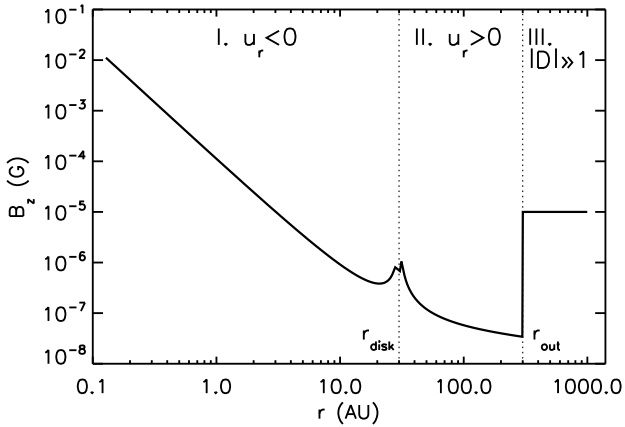}
\caption{Quasi-steady profile of the magnetic field, $B_{stdy}$,
 calculated by Equation (\ref{eq:B_stdy}) for the gas disk shown in
 Figure \ref{fig:gas_similar}. We set $P_{m,eff} C_u  =100$ for $r <
 r_{out}$. The spurious feature at $r_{disk}$ is an artifact of the
 approximate expression (see text). The disk is divided at $r_{disk}$
 and $r_{out}$.} 
\label{fig:Bz_stdy}
\end{figure}

The disk is divided into three parts: I. the inner accreting region,
II. the outer region where the gas moves outward, and III. the outermost
part where the gas cannot drag the magnetic field. These parts are
characterized as, 
\begin{equation}
\left\{ {\begin{array}{*{20}{c}}
   {{\rm{I}}{\rm{.}}} & {\bar{u_r}  < 0} & {{\rm{for}}} & {r <
    {r_{disk}}}  \\ 
   {{\rm{II}}{\rm{.}}} & {\bar{u_r}  > 0} & {{\rm{for}}} &
    {{r_{disk}} < r < {r_{out}}}  \\ 
   {{\rm{III}}{\rm{.}}} & {\left| D \right| \gg 1} & {{\rm{for}}} & {r >
    {r_{out}}}  \\ 
\end{array}} \right. \ ,
\label{eq:disk_part}
\end{equation}
as shown in Figure \ref{fig:Bz_stdy}.

\subsection{Approximate Steady Profile of the Magnetic Field for Strong Dragging}
\label{sec:B_stdy}

Steady solutions of the magnetic field have been derived in Paper I,
assuming that advection of the field is inward everywhere. As described
in Section \ref{sec:disk_struc}, we consider disks with both outward and
inward advections. We derive in Appendix \ref{sec:B_outward} an
approximate formula for the field strength in such disks.
Assuming that the field dragging is strong in the main body of the
disk ($|D| \ll 1$ for $r < r_{out}$), the vertical field strength is
approximately written as 
\begin{equation}
B_{stdy} = \left\{ {\begin{array}{*{20}{c}}
   2 \gamma_c {\left| D \right|\left( {\frac{{{r_{disk}}}}{{{r_{out}}}}}
		    \right){{\left( {\frac{r}{{{r_{disk}}}}} \right)}^{
		    - 2}}{B_\infty }} & {{\rm{for}}} & {r < {r_{disk}}}
		    \\ 
   {\frac{{2\left| D \right|}}{\pi}\left( {\frac{r}{{{r_{out}}}}}
				 \right){B_\infty }} & {{\rm{for}}} &
   {{r_{disk}} < r < {r_{out}}}  \\ 
   {{B_\infty }} & {{\rm{for}}} & {r > {r_{out}}}  \\
\end{array}} \right. \ ,
\label{eq:B_stdy}
\end{equation}
where the numerical factor $\gamma_c=0.43$. 

An example of the vertical field strength is shown in Figure
\ref{fig:Bz_stdy}. We set $P_{m,eff} C_u=100$ for $r < r_{out}$
($\tau_{dif} / \tau_{adv} = 100 \varepsilon$), making $|D| \approx
0.2-0.3$ for $r<r_{disk}$ and $|D| \sim 0.1-0.01$ for $r_{disk} < r <
r_{out}$. For $r > r_{out}$, we assume $|D| \gg 1$, and thus
$B_{stdy}=B_{\infty}$. 
The profile reflects the $r$ dependence of $|D|$. 
The magnetic field profile shown in Figure \ref{fig:Bz_stdy}
decreases with $r$ even in region II. It seems to be at odds with
Equation (\ref{eq:B_stdy}) and Figure \ref{fig:B_inout}. This behavior
of $B_{stdy}$ comes from the fact that $|D|$ is a decreasing function of
$r$. In the region II (and for $r \gg r_{disk}$), $\bar{\nu} \propto r$,
$\bar{u_r} \propto r$, $\varepsilon \propto r^{1/4}$, and then $|D| \propto
r^{-5/4}$. Further, $|D|$ becomes greater than unity when approaching
$r_{disk}$, because 
$\bar{u_r}(r_{disk})=0$.  Equation (\ref{eq:B_stdy}) is appropriate
only for $|D| \ll 1$. In Figure \ref{fig:Bz_stdy}, $|D|$ in Equation
(\ref{eq:B_stdy}) was replaced by $\min (|D|,1)$, which causes a
spurious feature of the field profile around $r_{disk}$. This feature is
actually not reproduced by the numerical calculation shown in Section
\ref{sec:Bz_evol}. We see that the magnetic field profiles in Figure
\ref{fig:Bevol_vi}-\ref{fig:Bevol_P1D1ec} are smooth around $r_{disk}$, 
and thus the spurious feature in the approximate expression should be 
considered as artificial.

Note that at the inner disk ($r < r_{disk}$) where the magnetic
flux moves inward, the field profile approaches $B_{stdy} \propto
r^{-2}$ for $|D| \rightarrow 0$
\footnote{We neglect in this paragraph $r$ dependence of $D$. In our
model, $D \propto \varepsilon^{-1}  \propto r^{-0.25}$, but this weak
dependence does not change the conclusion of this paragraph that $|D|
\ll 1$ is required for the magnetic flux to be tightly bundled.}
as shown in Figure 8 of Paper I and in Figure 2 of Guilet \& Ogilvie
(2014). This means that for $|D| \ll 1$ the magnetic flux, $\psi = \int
r B_z dr \propto \log r$, concentrates at the innermost part of the
disk. For weaker field dragging ($|D| \sim 1$; see section
\ref{sec:caseC} and Figure \ref{fig:Bevol_P1D1ec} below), though the magnetic
field $B_{stdy}$ looks to be still advected to the inner part,
concentration of the magnetic flux is more milder (see also Paper I; Guilet \& Ogilvie
2014), and consequently the magnetic flux $\psi (r)$ extends in the whole disk. In
this paper we consider that tight bundling of the magnetic flux occurs when $|D| \ll 1$.

\section{Time Evolution of the Magnetic Field}
\label{sec:Bz_evol}

In this section, we discuss the time evolution of the disk gas and the
magnetic field. Induction Equation (\ref{eq:induction_ave}) is
numerically solved under the evolution of the gas disk described by the
similarity solution (\ref{eq:ur_similar}). Note that only the velocity
profile $\bar{u_r}(r)$ is needed. 

The numerical method solving Equation(\ref{eq:induction_ave}) is similar
to that by LPP94. The computational domain is $[0.1{\rm AU},10^3{\rm
AU}]$, and 400 grid points are used with spacing in proportion to $r^{1/2}$. The disk region III where $|D| \gg 1$ is realized by setting
$C_u=0.1$ (Case A; Section \ref{sec:caseA}) or $P_{m,eff}=0.1$ (Case B
and C; Section \ref{sec:caseB} and \ref{sec:caseC}) for $r >
r_{out}$. For numerical stability, we further set $u_{r*}=0$ for
$r>800$AU. 

The initial profile of the gas disk is shown in Figure
\ref{fig:gas_similar}. We consider two types of initial profiles with
different vertical field strength. The first one is the uniform external field
$B_{\infty}$. This represents an extreme case in which magnetic flux
concentration has not occurred in the disk formation phase. The second
one represents a strong initial concentration of the magnetic flux. In
Paper I, we have discussed that the maximum field strength in a
steady state is expected to be 
\begin{equation}
 {B_{\max }} = \left\{ {\begin{array}{*{20}{c}}
   {{{\left( {\frac{r}{r_{out}}} \right)}^{ - 2}}{B_\infty }} & {{\rm{for}}} & {r < {r_{out}}}  \\
   {{B_\infty }} & {{\rm{for}}} & {r > {r_{out}}}  \\
\end{array}} \right.
\label{eq:Bmax}
\end{equation}
We set $B_z (t=0)=B_{\max}$ for the second extreme case. 

\begin{figure}[ptb]
\epsscale{1.2}
\plotone{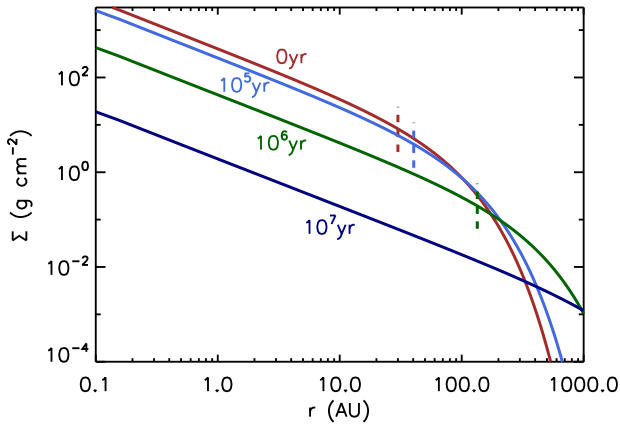}
\caption{Evolution of the gas density profile. The gas evolves
 self-similarly. The locations where the gas velocity switches
 from inward to outward are shown by the dashed vertical lines
 plotted on the density profiles.} 
\label{fig:gas_evol}
\end{figure}

The evolution of the gas disk is shown in Figure \ref{fig:gas_evol}. The
evolution timescale at the initial state is $\tau_{disk} \sim r_0^2 /
\bar{\nu_0} \sim 4 \times 10^5$yr, where disk radius $r_0=30$AU at 
$t=0$. As the disk evolves, its evolution timescale increases, because
the disk radius expands. In Figure \ref{fig:gas_evol} the locations of
$r_{disk}$ are shown by the dashed lines marked on the density profiles. It
is apparent that $r_{disk}$ moves outward with time.

\subsection{Case A: $\tau_{dif}=\varepsilon \tau_{disk}$ and
  $\tau_{adv}=10^{-2} \tau_{disk}$ ($C_u = 100$ and $P_{m,eff}=1$)}
\label{sec:caseA}

\begin{figure}[ptb]
\epsscale{1.2}
\plotone{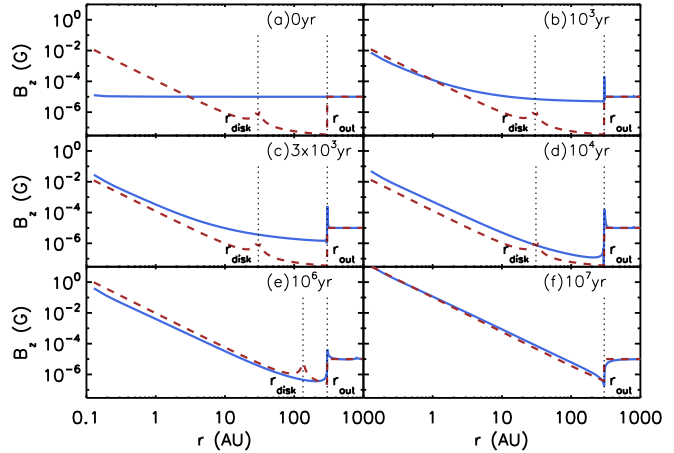}
\caption{Evolution of the magnetic field, $B_z$, for the case
 where $C_u=100$, $P_{m,eff}=1$ ($\tau_{dif}=\varepsilon \tau_{disk}$,
 $\tau_{adv}=10^{-2} \tau_{disk}$), and the initial profile of 
 $B_z=B_{\infty}$. The blue solid line shows the result of numerical
 integration. The brown dashed line shows the approximate steady
 field calculated by Equation (\ref{eq:B_stdy}) for the gas disk profile
 at each time. The vertical dotted lines show the radii $r_{disk}$ and
 $r_{out}$.} 
\label{fig:Bevol_vi}
\end{figure}

First, we consider the case with $C_u = 100$ and $P_{m,eff}=1$. This
corresponds to region A in Figure \ref{fig:Pm-Cu}. 
Figure \ref{fig:Bevol_vi} shows the evolution of the initially uniform
field $B_z=B_{\infty}$ at $t=0$. 
The magnetic flux is advected inward for $r < r_{disk}$ and outward
for $r > r_{disk}$. Because the advection timescale is shorter for smaller $r$,
the field profile approaches the quasi-steady state from inside (Figure
\ref{fig:Bevol_vi}b), and the part inside $r_{disk}$ becomes nearly
quasi-steady state at $t = \tau_{adv} = C_u^{-1} \tau_{disk}
\approx 4 \times 10^3$ yr (Figure \ref{fig:Bevol_vi}c). Then, the
magnetic flux outside $r_{disk}$ is expelled due to outward gas
migration. The outward advection velocity of the field is $u_{r*}
\propto r$ (Equation (\ref{eq:ur_similar}) with $\gamma = 1$), and 
the advection timescale outside $r_{disk}$ is nearly independent of $r$,
i.e., $r / u_{r*} \approx r_{disk} / u_{r*} (r_{disk}) =
\tau_{adv}$. Thus, the field profile outside $r_{disk}$ quickly evolves
with the timescale at $r_{disk}$ even for larger $r$. It takes several
times more than $\tau_{adv}$ for relaxation to the quasi-steady state in
the whole disk, because the initial field profile is quite different from the quasi-steady
field by orders of magnitude. The whole field profile becomes almost
the quasi-steady state in $t = 10^4$ yr (Figure \ref{fig:Bevol_vi}d),
which is much earlier than $\tau_{disk}(t=0) \sim 4 \times 10^5$yr, where we used
$\varepsilon=7.7 \times 10^{-2}$ at $r_0=30$AU. After field
relaxation, the magnetic field evolves as such that its profile keeps
the quasi-steady profile $B_{stdy}$, as seen in Figure
\ref{fig:Bevol_vi}e. Finally, in Figure \ref{fig:Bevol_vi}f at 
$10^7$yr, the disk radius $r_{disk}$ has expanded beyond $r_{out}$,
outside which the gas cannot drag the magnetic field. The magnetic field
shown in Figure \ref{fig:Bevol_vi}f is the final profile without
further evolution, provided that $r_{out}$ is fixed.  The evolutional
sequence shown in Figure \ref{fig:Bevol_vi} is controlled mainly by
advection. This is consistent with the analytical picture described in
Figure 4 of Guilet \& Ogilvie (2014). 

\begin{figure}[ptb]
\epsscale{1.2}
\plotone{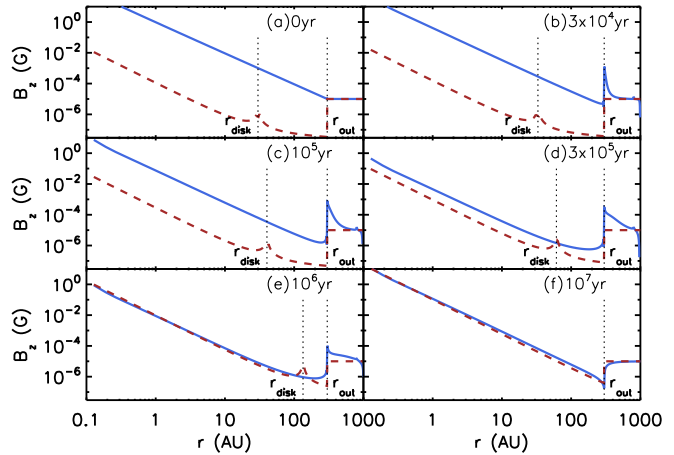}
\caption{Evolution of the magnetic field, $B_z$, for the case
 where $C_u=100$, $P_{m,eff}=1$ ($\tau_{dif}=\varepsilon \tau_{disk}$,
 $\tau_{adv}=10^{-2} \tau_{disk}$), and the initial profile of
 $B_z=B_{\max}$. The blue solid line shows the result of numerical
 integration and the brown dashed line shows the approximate steady
 solution.} 
\label{fig:Bevol_vc}
\end{figure}

Figure \ref{fig:Bevol_vc} shows the evolution of the magnetic field
which is concentrated to the center at $t=0$. The initial profile
$B_z=B_{\max} \propto r^{-2}$ is given by Equation (\ref{eq:Bmax}).
The magnetic flux is confined to the
innermost part of the disk ($\psi(r) \propto \log r$). In this case, the
evolution is controlled mainly by outward diffusion of the flux to
$r_{disk}$. The flux outside $r_{disk}$ is quickly advected to $r_{out}$
by the outward migrating gas with timescale $\tau_{adv} (< \tau_{disk})$, as
discussed in the last paragraph. Actually, in the numerical solution, 
evolution due to magnetic diffusion appears at $t =3 \times 10^4 \approx
\tau_{dif}$ (Figure \ref{fig:Bevol_vc}b),  and the magnetic field
profiles has nearly relaxed to the quasi-steady state by $t =3 \times
10^5 \approx 10 \tau_{dif} (t=0)$ (Figure \ref{fig:Bevol_vc}d). It takes about
$10 \tau_{dif} (t=0)$ for relaxation, because the initial field is about
$10^3$ times stronger than the quasi-steady field (compare the 
solid line in Figure \ref{fig:Bevol_vc}a and the dashed line in Figure
\ref{fig:Bevol_vc}d), and also because $\tau_{dif}$ increases as
$r_{disk}$ moves to 60 AU by $3 \times 10^5$ yr. Further evolution is
similar to the case of the initially uniform field in Figure
\ref{fig:Bevol_vi}, i.e., the magnetic field is in the quasi-steady
state (Figures \ref{fig:Bevol_vc}e and f).

The above numerical experiments with two extreme initial
conditions show that, if the effective magnetic Prandtl number
$P_{m,eff}$ is of order unity, the magnetic field quickly relaxes to 
the quasi-steady profile $B_{stdy}$. 
The relaxation timescale is several times the characteristic
timescale either of advection, $\tau_{adv}=C_u^{-1} \tau_{disk}$, or of
diffusion, $\tau_{dif} = \varepsilon P_{m,eff} \tau_{disk}$, depending on the
initial field profile. It is shorter than the gas evolution
timescale unless the initial magnetic flux in the disk is too strong.
Thus, except for a beginning stage of the gas evolution,
the steady solution derived in Paper I gives plausible estimates for
the magnetic field strength of actual disks.

\subsection{Case B: $\tau_{dif}=100 \varepsilon \tau_{disk}$ and
  $\tau_{adv}=\tau_{disk}$ ($C_u = 1$ and $P_{m,eff}=100$)}
\label{sec:caseB}

\begin{figure}[ptb]
\epsscale{1.2}
\plotone{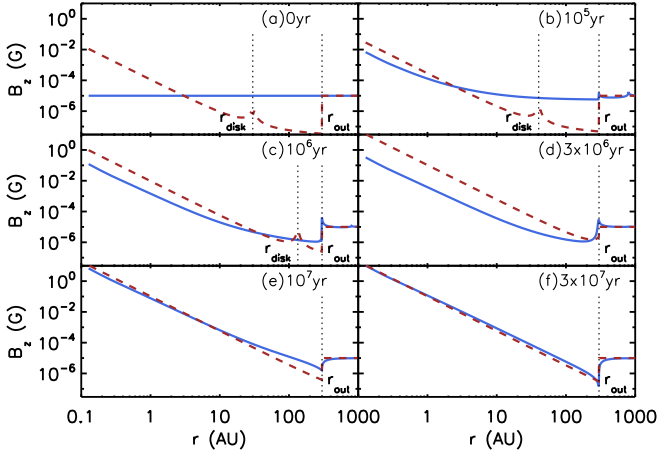}
\caption{Evolution of the magnetic field, $B_z$, for the case
 where $C_u=1$, $P_{m,eff}= 100$ ($\tau_{dif}=100 \varepsilon
 \tau_{disk}$, $\tau_{adv}=\tau_{disk}$), and the initial profile of 
 $B_z=B_{\infty}$. The blue solid line shows the result of numerical
 integration and the brown dashed line shows the approximate steady
 solution.} 
\label{fig:Bevol_ei}
\end{figure}

\begin{figure}[ptb]
\epsscale{1.2}
\plotone{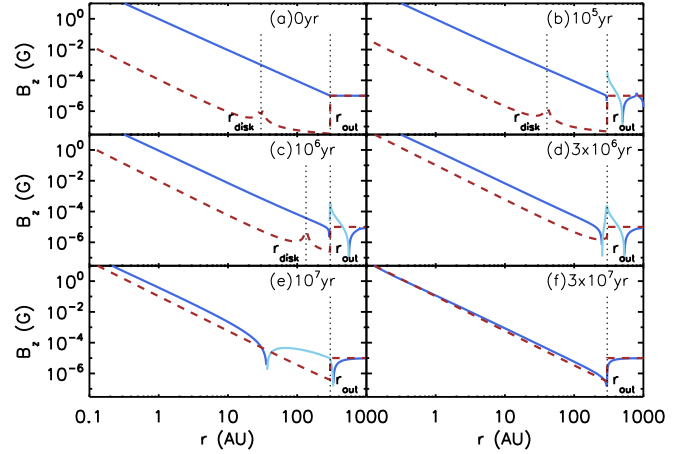}
\caption{Evolution of the magnetic field, $B_z$, for the case
 where $C_u=1$, $P_{m,eff}= 100$ ($\tau_{dif}=100 \varepsilon
 \tau_{disk}$, $\tau_{adv}=\tau_{disk}$), and the initial profile of
 $B_z=B_{\max}$. The blue solid line shows the result of numerical
 integration and the brown dashed line shows the approximate steady
 solution. The part shown by the light blue line has the negative values of
 $B_z$.} 
\label{fig:Bevol_ec}
\end{figure}

In this section, we consider the case in which $C_u = 1$ and $P_{m,eff} =
100$, which corresponds to region B in Figure \ref{fig:Pm-Cu}. Inward
field advection is maintained by large effective Prandtl number $P_{m,eff}$, not
by high advection speed of the magnetic field ($C_u = 1$). Under the
above parameters, the 
diffusion timescale of the magnetic field is $\tau_{dif} \sim P_{m,eff}
\varepsilon \tau_{disk} \sim 3 \times 10^6$yr, which is much slower than
the disk evolution timescale $\tau_{disk}$. 
Figures \ref{fig:Bevol_ei} and \ref{fig:Bevol_ec} show that the magnetic
field remembers the initial profile more than $10^7$yr. The field
relaxes to the steady state after $3 \times 10^7$yr, but at that time the
disk has already experienced considerable evolution. These calculations
have shown that the steady solution derived in Paper I would not be
appropriate in Case B.  

The numerical calculations sometimes result in negative values of $B_z$
around $r_{out}$, that are shown by the light-blue lines in Figures 
\ref{fig:Bevol_ec}b$-$e. 
This negative $B_z$ originates from our simplified
formalism of the field evolution and the initial condition adopted
here. At the initial state shown in Figure \ref{fig:Bevol_ec}a, the magnetic field is
much stronger than that expected as the quasi-steady state (the dashed
line). The magnetic flux inside $r_{out}$ is also much stronger than the
analytical upper limit $2 \pi r_{out}^2 B_{\infty}$ derived by Okuzumi et
al. (2014). Such a strong concentration of the magnetic flux at the initial state is
maintained not only by the toroidal current inside $r_{out}$ but also by
the current outside $r_{out}$. Because we
assume $P_{m,eff}=0.1$ outside $r_{out}$, the current there quickly
dissipates, while the current inside $r_{out}$ remains for much longer
time. The remaining current especially at $r_{out}$ generates
negative $B_z$. (Negative $B_z$ appears even when we set $P_{m,eff} =1$
outside $r_{out}$.) Note that appearance of negative 
$B_z$ means formation of closed field lines connecting different radii
in the differentially rotating disk. They should give rise to toroidal fields
that are ignored in our model, suggesting that field evolution in Figure
\ref{fig:Bevol_ec} is not self-consistent. The negative $B_z$ disappears
by $3 \times 10^7$ yr and the quasi-steady state is finally realized (Figure
\ref{fig:Bevol_ec}f). Because this relaxation timescale is similar to
that with the initially uniform field in Figure \ref{fig:Bevol_ei}, we
expect our qualitative conclusion on the relaxation timescale is
correct, but it should be examined by a more sophisticated model.

\subsection{Case C: $\tau_{dif}=10 \varepsilon \tau_{disk}$ and
  $\tau_{adv}=\tau_{disk}$ ($C_u = 1$ and $P_{m,eff}=10$)}
\label{sec:caseC}

\begin{figure}[ptb]
\epsscale{1.2}
\plotone{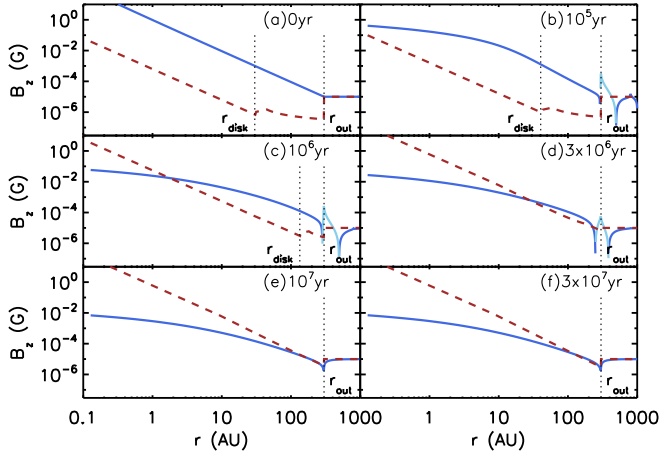}
\caption{Evolution of the magnetic field, $B_z$, for the case
 where $C_u=1$, $P_{m,eff}= 10$ ($\tau_{dif}=10 \varepsilon
 \tau_{disk}$, $\tau_{adv}=\tau_{disk}$), and the initial profile of 
 $B_z=B_{\max}$. The blue solid line shows the result of numerical
 integration and the brown dashed line shows the approximate steady
 solution. The part shown by the light blue line has the negative values of
 $B_z$.} 
\label{fig:Bevol_P1D1ec}
\end{figure}

If $C_u P_{m,eff} < \varepsilon^{-1}$ ($|D| \ga 1$), the magnetic flux diffuses
outward more quickly compared to the inward field advection. Figure
\ref{fig:Bevol_P1D1ec} shows the magnetic field evolution with $C_u = 1$
and $P_{m,eff}=10$. Initial field profile $B_z(t=0) = B_{\max}$ shows
concentration of the flux toward $r=0$. This flux that initially
concentrated to the center diffuses outward, and at $10^7$yr the profile
reaches a steady state, which shows only weak concentration of the flux
at the center. The brown line in Figure \ref{fig:Bevol_P1D1ec} 
shows the quasi-steady profile given by Equation (\ref{eq:B_stdy}),
which assumes $|D| \ll 1$. Here, actual $|D| \sim (P_{m,eff} C_u
\varepsilon)^{-1} \sim 1$, and the resultant magnetic field is much
weaker than the field strength expected for cases A and B with strong
inward dragging. This result confirms the conclusion of LPP94 that strong field
dragging (or flux concentration to the inner part) requires $|D| \ll 1$.

\section{Discussions}
\label{sec:Discussions}

\subsection{Dependence of $r_{out}$}
\label{sec:Dis_rout}

We used a fixed value $r_{out}=300$AU for simplicity in the numerical
calculations in Section \ref{sec:Bz_evol}. The steady
profile described in Section \ref{sec:B_stdy} depends on $r_{out}$ 
(Equation(\ref{eq:B_stdy})). Thus, $r_{out}$ is an important parameter
for determining the magnetic field strength in (quasi-)steady states.
In real disks, $r_{out}$ should be determined as the radius outside which the
gas can no longer drag the field effectively. Determining $r_{out}$
requires further investigations on the tenuous outermost part of disks.

The evolutionary sequences described in Section \ref{sec:Bz_evol} do not
depend on the value of $r_{out}$. Time evolutions of the magnetic field
are classified into three regimes A-C in Figure \ref{fig:Pm-Cu}, based
on $P_{m,eff}$ and $C_u$. Each regime is characterized via
comparison of the several evolution timescales, which are evaluated at
$r_{disk}$, not at $r_{out}$. Thus, $r_{out}$ does not affect the
classification in Figure \ref{fig:Pm-Cu}. To see this, we have performed
additional calculations with a different assumption on $r_{out}$. In
those additional calculations, we 
assume that the boundary between the regions II and III in Figure
\ref{fig:Bz_stdy} is determined by the density, i.e., $r_{out}$ is
calculated from the condition $\Sigma(r_{out})=\Sigma_c$, where
$\Sigma_c$ is the threshold value. As $\Sigma$ evolves with time,
$r_{out}$ varies. We performed calculations for $\Sigma_c=10^{-2}$,
$0.1$, and $1$, and confirmed that the classification of the
evolutionary sequences does not depend on $r_{out}$. 

\subsection{Vertical Structure of the Disk}

In Section \ref{sec:timescales}, the advection and diffusion timescales
were discussed. Estimate of the timescales is based on induction
equation (\ref{eq:induction_ave}), which is vertically averaged with
conductivity weighting. (We call $1/\eta$ as conductivity.) The
differences between the conductivity-weighted 
and mass-weighted averages are expressed by the parameters $P_{m,eff}$
and $C_u$ given by Equations (\ref{eq:ur_ave_ratio}) and
(\ref{eq:Pmeff}). In this paper, we treat $P_{m,eff}$
and $C_u$ as free parameters and see how the magnetic field evolution is
controlled by these parameters. 
In reality, averaging requires knowledge of the vertical structure
of the disk, such as $\eta(z)$, $\nu(z)$, and $u_r (z)$.

In protoplanetary disks, the ionization degree decreases with the depth
from the disk surface (Sano et al. 2000), and thus the Ohmic resistivity
increases with the depth. In disks with MRI turbulence, turbulent
resistivity would be added. The MRI is active around a certain 
height which is above the dead zone around the midplane (if present) and
which is below the disk corona. At the disk corona, ambipolar diffusion is
responsible for the magnetic diffusivity. To determine the height
dependence of the magnetic diffusivity, contribution of Ohmic,
turbulent, and ambipolar diffusivities should be calculated. There have
been calculations of Ohmic and ambipolar diffusivities in laminar
stratified disks (Walsh et al. 2012; Dzyurkevich et al. 2013), and also
numerical simulations of measuring the turbulent diffusivity in a local
shearing box (Guan \& Gammie 2009; Lesur \& Longaretti 2009; Fromang \&
Stone 2009). We need to combine these results and measure the total
diffusivity in the turbulent stratified disk by numerical simulations.

Vertical profile of the gas velocity $u_r (z)$ is also important, because
the conductivity-weighted average $u_{r*}$ is dominated by $u_r (z)$ at
specific regions where $\eta (z)$ has a minimum value which
possibly occurs at the surface of the disk. If $|u_r|$ is an increasing function
of $|z|$, $C_u = u_{r*} / \bar{u_r}$ could be much larger than unity and
also would be larger for higher $\beta_z$ (see the lower left panel of Figure 8 in Guilet \& Ogilvie
2012).  If $C_u$ is expressed by a function of $\beta_z$ and $P_{m,eff}$,
then Figure \ref{fig:Pm-Cu} can be converted on the $P_{m,eff}$-$\beta_z$
plane. As an example, using the results shown in Figure 12 of Guilet \& Ogilvie
(2012), the boundary between the hatched region and region C
($\tau_{adv}=\tau{dif}$) corresponds to $\beta_z=10^3$ for $P_{m,eff}=1$
and $\beta_z=1$ for $P_{m,eff}=\varepsilon^{-1}$. Note that these values
could be quite different for other disk models, depending on the
vertical structure of the disk.

\section{Summary}
\label{sec:summary}

We have studied the time evolution of a large-scale magnetic flux threading
an accretion disk. Induction equation of the mean poloidal field is
solved for a viscously evolving disks.

We use mass-weighted averaging for the equation of surface density
evolution, while conductivity-weighted averaging is used for the
induction equation, according to suggestion by Ogilvie \& Livio (2001).
In a thin disk approximation, fluid equations and induction equation are
averaged in the vertical direction. While mass-weighted averaging method
is useful for gas dynamics, conductivity-weighted averaging method is
more appropriate for magnetic flux evolution. There may be differences
between these two methods. The ratios between the mass-weighted and conductivity
weighted averages ($C_u$ and $P_{m,eff}$ defined in Equations
(\ref{eq:ur_ave_ratio}) and (\ref{eq:Pmeff})) are the important
parameters determining the evolution timescales of the gas and magnetic
flux.

The ratio of the magnetic field evolution timescale to the disk
evolution timescale is controlled by the parameter $C_u$ and
$P_{m,eff}$. The diffusion timescale of the magnetic flux $\tau_{dif}$
is $\varepsilon P_{m,eff}$ times the disk evolution timescale,
$\tau_{disk}$, where $\varepsilon \ll 1$ is the geometric aspect ratio of the
disk ($\tau_{dif} = \varepsilon P_{m,eff} \tau_{disk}$). 
The advection timescale of the magnetic flux is written as
$\tau_{adv}=C_u^{-1} \tau_{disk}$. The evolution of the magnetic flux
can be categorized by these timescales (or the parameters $P_{m,eff}$ and
$C_u^{-1}$).

Using these timescales we categorize evolutional types of magnetic fields.
If $\tau_{dif} \ll \tau_{disk}$ (or $P_{m,eff} \la \varepsilon^{-1}$), the magnetic flux quickly relaxes to a quasi-steady profile. Thus,
the field profile in each evolutionary phase of the gas disk is given by
the steady profiles discussed in Paper I. Further, if $\tau_{adv} \ll
\tau_{disk}$ (or $C_u \gg 1$) is satisfied at the same time, the magnetic flux
profiles would have relaxed to a quasi-steady state in which the flux
would be tightly bundled at the inner part of the disk. This regime of
the magnetic flux evolution is shown in Figure \ref{fig:Pm-Cu} as the region A.
On the other hand, even if $\tau_{adv} \ll \tau_{disk}$, in the case of
$\tau_{dif} \ga \tau_{disk}$ (or $P_{m,eff} \ga \varepsilon^{-1}$), the initial profile of
the magnetic flux remains longer than the disk evolution timescale.
In this case, the initial disk formation phase is important for determining
later magnetic flux evolution. This evolutionary regime is the region B
in Figure \ref{fig:Pm-Cu}. 
Finally, if $\tau_{dif} \ll \tau_{adv}$ (or $C_u P_{m,eff} <
\varepsilon^{-1}$), the accreting gas cannot drag the magnetic flux
significantly, and only weak concentration of the 
magnetic flux in the disk is expected. This regime is shown as the
region C in Figure \ref{fig:Pm-Cu}.

This paper treats $P_{m,eff}$ and $C_u$ as free parameters, and we do not
specify the timescales $\tau_{dif}$ and $\tau_{adv}$. In reality, these
should be physically determined via the vertical structure of the
disk and field lines. The key issues are the height where the
magnetic field dragging occurs most effectively (Ogilvie \& Livio 2001;
Rothstein et al. 2008) and the angular momentum extract by the disk wind
(Bisnovatyi-Kogan \& Lovelace (2012); Guilet \& Ogilvie (2012)).
The efficiency of field dragging should be determined by future
investigations on vertical dependences of $u_r (z)$, $\nu (z)$,
and $\eta (z)$ and structure of field lines. How much the disk wind
would extract the angular momentum from the disk also should be quantified.

Finally, we should note that this paper is based on the LPP94 model in
which any effect due to toroidal fields is ignored. Toroidal fields transfer angular
momentum along {\boldmath $B$}, and possibly launch disk wind, which
extract the angular momentum from the disk. Global evolution of magnetic
fields for wind driven accreting disks is an important topic. Further,
presence of the disk wind may cause a toroidal current above the
Alfv\'en surface. Ogilvie (1997) discussed that the effect of the
toroidal current outside the Alfv\'en surface effectively works as modifying the external
field {\boldmath $B$}$_{\infty}$, which is the boundary condition in
the LPP94 model at infinity. With this modification, the field
configuration and the flux transport inside the Alfv\'en surface can be
treated by the LPP94 model. However, how much {\boldmath $B$}$_{\infty}$
is modified by the wind has not been clear and needs to be quantified in
future works. 

\acknowledgements We thank Takayuki Muto for useful discussions and
Junko Kominami for careful reading of the manuscript. 
We appreciate the anonymous referee for his/her thorough review that
helped us to improve the quality of the paper and to find errors in the
original version of the manuscript.
This work was supported by Grants-in-Aid for Scientific Research, Nos. 20540232,
23103005, 25887023, 26103704, and 26400224 from MEXT of Japan.

\appendix

\section{Steady Profile of the Magnetic Field under Outward Advection}
\label{sec:B_outward}

We have derived in Paper I the approximate expressions of the steady
magnetic profiles in accretion disks, assuming that advection of the
magnetic field is inward everywhere in the disk. In reality, the advection is
outward in the outer part of disks. In this Appendix, we derive the 
approximate expressions of the magnetic field profiles in disks with outward
motions.

The vertically averaged induction equation is
\begin{equation}
B_z = \frac{D}{2} \frac{{4\pi {K_{\phi}}}}{c} \ ,
\label{eq:induction_stdy2}
\end{equation}
where $D$ is negative for outward advection ($u_{r*}>0$). We first consider
a piecewise profile of $D$ as
\begin{equation}
D = \left\{ {\begin{array}{*{20}{c}}
   {{D_{II}}} & {{\rm{for}} \ \ \ {r_{in}} < r < {r_{out}}}  \\
   \infty  & {{\rm{otherwise}}}  \\
\end{array}} \right. \ ,
\end{equation} 
where $D_{II}$ is negative and $|D_{II}| \ll 1$. 

\begin{figure}[ptb]
\epsscale{0.55}
\plotone{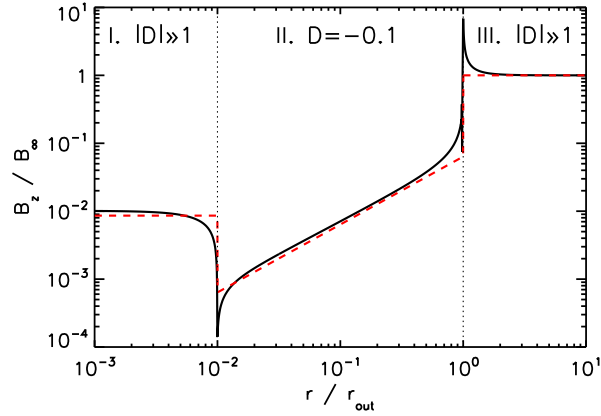}
\caption{Steady profile of the magnetic field for piecewise constant
 $D$. The solid line is numerically calculated, and the dashed line
 shows the approximate expression given by Equation (\ref{eq:Bz_apx}).} 
\label{fig:B_outward}
\end{figure}

The numerical solution of Equation (\ref{eq:induction_stdy2}) is shown
in Figure \ref{fig:B_outward}. We set $r_{in}=10^{-2}$, $r_{out}=1$, and
$D_{II}=-0.1$. From Figure \ref{fig:B_outward}, it is apparent that $B_z \ll
B_{\infty}$ for $r < r_{out}$. Thus, most of the imposed external flux,
$\psi_{\infty}=B_{\infty} r^2 /2$, has been expelled from the disk. The
resultant magnetic flux $\psi$ is almost zero for $r< r_{out}$ compared
to $\psi_{\infty}$. For $r_{in} < r < r_{out}$, the vertical field
strength is proportional to $r$, in contrast to the inward advection
case (positive $D$) where $B_z \propto r^{-2}$. 

The profile of the magnetic field is approximately expressed by
\begin{equation}
{B_z} = \left\{ {\begin{array}{*{20}{c}}
   {2\gamma_c \left( {\frac{{{r_{in}}}}{{{r_{out}}}}} \right){B_\infty }} & {{\rm{for}}} & {r < {r_{in}}}  \\
   {\frac{{2\left| D \right|}}{\pi }\left( {\frac{r}{{{r_{out}}}}} \right){B_\infty }} & {{\rm{for}}} & {{r_{in}} < r < {r_{out}}}  \\
   {{B_\infty }} & {{\rm{for}}} & {r > {r_{out}}}  \\
\end{array}} \right. \ ,
\label{eq:Bz_apx}
\end{equation}
which is shown by the red dashed line in Figure \ref{fig:B_outward}. In
Equation (\ref{eq:Bz_apx}), $\gamma_c=0.43$ is a numerical constant 
which was derived in Paper I. 

In the following subsections, we derive this approximate
solution. According to the procedure described in Appendix B of Paper I,
the profile of the surface current is expanded in a power-law series in
$r$. Then, Biot-Savart Equation (\ref{eq:BiotSavart}) is integrated to
obtain the disk potential $\psi_d$. The disk potential should be in a
quadratic form of $r$, which determines the profile of the surface current.

\subsection{Solutions near the Outer Boundary}
Following Paper I, the surface current near the outer boundary $r_{out}$
is expanded as
\begin{equation}
K_{\phi}(r') = A_{out}r{'^{ - 2}}\left[ {{a_0} +
 {{\sum\limits_{m=0}^\infty {{a_{2m+1}} \left(
 {\frac{{r'}}{{{r_{out}}}}} \right)} }^{2m+1}}} \right] \ .
\label{eq:Js_out_expand}
\end{equation}
The expansion starts from $r^{-2}$. For positive $D$ cases described in
Paper I, it is expected that $K_{\phi} \propto r^{-2}$ for $r \ll
r_{out}$. However, as shown in Figure \ref{fig:B_outward}, the surface
current $K_{\phi} \propto B_z$ is proportional to $r$ if $D$ is
negative. We show below that $a_0=a_1=0$ and actually $K_{\phi}
\propto r$ for $r \ll r_{out}$. Integration of Biot-Savart Equation
(\ref{eq:BiotSavart}) gives the disk potential 
\begin{eqnarray}
\psi_d &=& \frac{{2\pi A_{out}{a_0}}}{c} + \frac{{2\pi
 A_{out}{a_1} r}}{{c{r_{out}}}} \nonumber \\
& & + \frac{{\pi A_{out}}}{c}\sum\limits_{n = 0}^\infty  {{c_n}{{\left(
    {\frac{r}{{{r_{out}}}}} \right)}^{2(n+1)}} \left[ { -
    \frac{{{a_0}}}{{2(n + 1)}} + \sum\limits_{m = 0}^\infty 
      {\frac{{{a_{2m + 1}}}}{{2(m - n) - 1}}} } \right]} \ ,
\label{eq:psi_d_out1}
\end{eqnarray}
where the Laplace coefficients $c_n=[(1/2)_n (3/2)_n] / [(1)_n (2)_n]$
and $(a)_n=a (a+1) \cdot \cdot \cdot (a+n-1)$. The disk potential should
have a quadratic form of $r$ as
\begin{equation}
\psi_d = \psi_{in} - \frac{1}{2} B_{\infty} r^2 \ .
\label{eq:psi_d_form}
\end{equation}
Further, as seen in Figure \ref{fig:B_outward}, $\psi_{in}$ must be
negligible compared to the external flux $\frac{1}{2} B_{\infty} r^2$. 
Equating Equation (\ref{eq:psi_d_out1}) and Equation (\ref{eq:psi_d_form})
with $\psi_{in}=0$ gives the conditions on the coefficients $a_{2m+3}$ as
\begin{equation}
a_0=a_1=0 \ ,
\label{eq:am_cond1}
\end{equation}
\begin{equation}
\sum\limits_{m = 1}^\infty  {\frac{{{a_{2m + 3}}}}{{2(m - n) - 1}} =
 \frac{{{a_3}}}{{2n + 1}}} \ \ \ {\rm for} \ \ \ n \ge 0 \ .
\label{eq:am_cond2}
\end{equation}
In deriving Equation (\ref{eq:am_cond2}), the indexes $m$ and $n$ were
replaced by $m-1$ and $n-1$, respectively. When Equations
(\ref{eq:am_cond1}) and (\ref{eq:am_cond2}) are satisfied, the disk
potential becomes
\begin{equation}
{\psi _d} = \frac{{\pi A_{out}}}{c}{\left( {\frac{r}{{{r_{out}}}}}
 \right)^2}\left[ {{a_3} + \sum\limits_{m=1}^\infty {\frac{{{a_{2m +
 3}}}}{{2m+1}}} } \right] .
\label{eq:psi_d_out2}
\end{equation}
Condition for the coefficients $a_{2m+3}$ in Equation
(\ref{eq:am_cond2}) is the same as the condition at the inner boundary for
$D>0$ cases treated in Paper I. Comparing Equation (\ref{eq:am_cond2})
to Equations (B23) and (B26) in Paper I, we find that
\begin{equation}
\frac{{{a_{2m + 3}}}}{{{a_3}}} = \frac{{{{( -
1)}^m}}}{{m!}}\frac{{\Gamma \left( {1/2} \right)}}{{\Gamma (1/2 - m)}}
\ ,
\end{equation}
where we used $\kappa=1/2$. Inserting this expression to Equation
(\ref{eq:psi_d_out2}) and using 
\begin{equation}
\sum\limits_{m = 1}^\infty  {\frac{{{a_{2m + 3}}/{a_3}}}{{2m + 1}} =
 \frac{\pi }{2} - 1} \ ,
\end{equation}
the disk potential reduces to
\begin{equation}
{\psi_d} = \frac{{{\pi ^2}A_{out}{a_3}}}{{2c}}{\left( {\frac{r}{{{r_{out}}}}}
  \right)^2} \ .
\label{eq:psi_d_rout}
\end{equation}
Equating Equations (\ref{eq:psi_d_form}) and (\ref{eq:psi_d_rout}), we
find $2 \pi A_{out} a_3 /c = 2 B_{\infty} r_{out}^2 /\pi$. From Equations
(\ref{eq:induction_stdy2}) and (\ref{eq:Js_out_expand}), the vertical
field strength for $r_{in} \ll r < r_{out}$ is 
\begin{equation}
{B_z} = \frac{2}{\pi }\left| D \right|{B_\infty }\left(
 {\frac{r}{{{r_{out}}}}} \right){\left[ {1 - {{\left(
  {\frac{r}{{{r_{out}}}}} \right)}^2}} \right]^{-1/2}} \ ,
\label{eq:Bz_apx_out}
\end{equation}
where we used $[1-(r'/r_{out})^2]^{-1/2}=1+\Sigma_{m=1}^{\infty}
(a_{2m+3}/a_3) (r'/r_{out})^{2 m} $ (see Equations (B21) and (B24) in
Paper I).

\subsection{Solutions near the Inner Boundary}
Near the inner boundary $r_{in}$, the surface current is expanded as
\begin{equation}
K_{\phi} (r') = A_{in}r'\left[ {1 + {b_2}{{\left(
   {\frac{{r'}}{{{r_{in}}}}} \right)}^{ - 2}} + {{\sum\limits_{m =
1}^\infty  {{b_{2m + 1}}\left( {\frac{{r'}}{{{r_{in}}}}} \right)} }^{ -
(2m + 1)}}} \right] \ ,
\end{equation}
where we used the fact that $K_{\phi} \propto r$ for $r \gg
r_{in}$, and $b_1=b_4=b_6=b_8= \cdot \cdot \cdot =0$ for convergence of
integration in Equation (\ref{eq:BiotSavart}). The disk potential is
\begin{eqnarray}
 {\psi _d} & = &
 \frac{\pi A_{in} r_{in}^3}{c} \left\{ 2{b_2} \left(
 \frac{r}{r_{in}} \right) + 2{b_3} - \sum\limits_{n = 0}^\infty
 c_n \left( \frac{r}{r_{in}} \right)^{-(2n + 1)} \right. \nonumber \\
& & \times \left. \left[ \frac{1}{2(n + 2)} + \frac{b_2}{2(n + 3)} +
      \sum\limits_{m = 1}^\infty  \frac{b_{2m + 1}}{2(n - m) + 3}
	   \right] 
\right\} \nonumber \\
 & & + \frac{\pi A_{in} r_{out}^3}{c} \left( \frac{r}{r_{out}} \right)^2 \ ,
\label{eq:psi_d_in1}
\end{eqnarray}
where we used $r_{out} \gg r$. This expression must be equal to Equation
(\ref{eq:psi_d_form}). Note that, as shown in Figure
\ref{fig:B_outward}, total magnetic flux can be treated as a constant
$\psi_{in}$ near the inner boundary $r_{in} \le r \ll |D|^{-1/3} r_{in}$. We
find
\begin{equation}
b_2=0 \ ,
\label{eq:bm_cond1}
\end{equation}
\begin{equation}
\sum\limits_{m = 1}^\infty  {\frac{{{b_{2m + 1}}}}{{2(m - n) - 1}} =
\frac{{{b_0}}}{{2(n + 1)}}}  
\ \ \ {\rm for} \ \ \ n \ge 1 \ .
\label{eq:bm_cond2}
\end{equation}
When deriving Equation (\ref{eq:bm_cond2}) the index $n$ was replaced by
$n-1$. The above conditions are the same as that of the outer boundary
cases for $D>0$ treated in Paper I. Comparing Equation
(\ref{eq:bm_cond2}) to Equations (B15) and (B17) in Paper I, we find that
\begin{equation}
 {b_{2m + 1}} = \frac{{{{( - 1)}^m}}}{{m!}}\frac{{\Gamma \left(
 {\gamma_c  + 1} \right)}}{{\Gamma (\gamma_c  + 1 - m)}} \ ,
\end{equation}
where $\gamma_c=0.43$. Equating Equation (\ref{eq:psi_d_in1})
and Equation (\ref{eq:psi_d_form}), and using $b_3=-\gamma_c$, the disk
potential becomes
\begin{equation}
{\psi _d} = \frac{{\gamma_c {B_\infty }{r_{in}}^3}}{{{r_{out}}}} -
\frac{{{B_\infty }{r^2}}}{2} \ .
\end{equation}
The total flux is 
\begin{equation}
\psi  = \gamma_c \left( {\frac{{{r_{in}}}}{{{r_{out}}}}} \right){B_\infty
 }{r_{in}}^2 \ .
\label{eq:psi_tot_in}
\end{equation}
Equation (\ref{eq:psi_tot_in}) shows that the total magnetic flux inside
$r_{in}$ is reduced by a factor of $2 \gamma_c (r_{in}/r_{out})$ from the
flux of the external field. Assuming constant field strength for $r <
r_{in}$, $B_z$ is approximately written as
\begin{equation}
B_z \approx 2 \gamma_c \left( \frac{r_{in}}{r_{out}} \right) B_{\infty} \
 .
\label{eq:Bz_in}
\end{equation}
Note that Equation (\ref{eq:Bz_in}) is a crude estimate because
$r$-dependence of $B_z$ near $r_{in}$ is neglected.

\subsection{Magnetic Field Profile When Both Inward and Outward Advections
  Exist}

\begin{figure}[ptb]
\epsscale{0.55}
\plotone{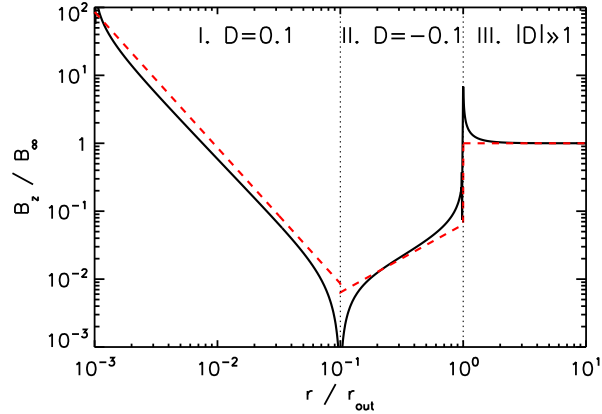}
\caption{Steady profile of the magnetic field for piecewise constant
 $D$.  The solid line is numerically calculated, and the dashed line
 shows the approximate expression given by Equation (\ref{eq:Bz_apxinout}).} 
\label{fig:B_inout}
\end{figure}

Using the results described above and in Paper I, we can construct
approximate expressions for magnetic field profiles in disks where both
inward and outward advections exist. We consider the case,
\begin{equation}
D = \left\{ {\begin{array}{*{20}{c}}
   D_{I}  > 0 & {{\rm{for}}} & {r < {r_{disk}}}  \\ 
   D_{II} < 0 & {{\rm{for}}} & {{r_{disk}} < r < {r_{out}}}  \\ 
   \infty & {{\rm{for}}} & {r >
    {r_{out}}}  \\ 
\end{array}} \right.
\label{eq:D_piece2}
\end{equation}
The magnetic field profile can be determined from the outside. For $r >
r_{out}$, the magnetic field is constant $B_{\infty}$. For $r_{disk} < r
< r_{out}$, the profile is given by Equation (\ref{eq:Bz_apx_out}). For
$r < r_{disk}$, the profile is given by Equation (40) of Paper I with
a modification that the total flux is reduced by a factor $2 \gamma_c
(r_{disk}/r_{out})$. Thus, $B_z$ is written as
\begin{equation}
{B_z} = \left\{ {\begin{array}{*{20}{c}}
   2 \gamma_c {\left| D \right| \left( {\frac{{{r_{disk}}}}{{{r_{out}}}}}
		    \right){{\left( {\frac{r}{{{r_{disk}}}}} \right)}^{
		    - 2}}{B_\infty }} & {{\rm{for}}} & {r < {r_{disk}}}
		    \\ 
   {\frac{{2\left| D \right|}}{\pi}\left( {\frac{r}{{{r_{out}}}}}
				 \right){B_\infty }} & {{\rm{for}}} &
   {{r_{disk}} < r < {r_{out}}}  \\ 
   {{B_\infty }} & {{\rm{for}}} & {r > {r_{out}}}  \\
\end{array}} \right. \ .
\label{eq:Bz_apxinout}
\end{equation}
This approximate expression is compared to the numerical result in
Figure \ref{fig:B_inout}. Some differences between the approximate
formula and the numerical result are apparent. For $r<r_{disk}$, 
Equation (\ref{eq:Bz_apxinout}) gives smaller $B_z$ than the numerical
result by a factor $0.64 \approx 2/\pi$, implying that a more rigorous
treatment at $r_{disk}$ is required. Approximate Equation
(\ref{eq:Bz_apxinout}) (and (\ref{eq:B_stdy}) in the main text) should
be considered to have errors by a factor of $\approx 2$.

\end{document}